\documentstyle[epsf]{mn}
\newif\ifAMStwofonts
\AMStwofontstrue

\ifoldfss
  \ifCUPmtlplainloaded \else
    \NewTextAlphabet{textbfit} {cmbxti10} {}
    \NewTextAlphabet{textbfss} {cmssbx10} {}
    \NewMathAlphabet{mathbfit} {cmbxti10} {} 
    \NewMathAlphabet{mathbfss} {cmssbx10} {} 
  \fi
  \ifAMStwofonts
    \ifCUPmtlplainloaded \else
      \NewSymbolFont{upmath} {eurm10}
      \NewSymbolFont{AMSa} {msam10}
      \NewMathSymbol{\upi}     {0}{upmath}{19}
      \NewMathSymbol{\umu}     {0}{upmath}{16}
      \NewMathSymbol{\upartial}{0}{upmath}{40}
      \NewMathSymbol{\leqslant}{3}{AMSa}{36}
      \NewMathSymbol{\geqslant}{3}{AMSa}{3E}

       \let\le=\leqslant
       
    \fi
  \fi
\fi 

\ifnfssone
  \newmathalphabet{\mathit}
  \addtoversion{normal}{\mathit}{cmr}{m}{it}
  \addtoversion{bold}{\mathit}{cmr}{bx}{it}
  \newmathalphabet{\mathbfit} 
  \addtoversion{normal}{\mathbfit}{cmr}{bx}{it}
  \addtoversion{bold}{\mathbfit}{cmr}{bx}{it}
  \newmathalphabet{\mathbfss} 
  \addtoversion{normal}{\mathbfss}{cmss}{bx}{n}
  \addtoversion{bold}{\mathbfss}{cmss}{bx}{n}
  \ifAMStwofonts
    \ifCUPmtlplainloaded \else
      \UseAMStwoboldmath
      \makeatletter
      \new@mathgroup\upmath@group
      \define@mathgroup\mv@normal\upmath@group{eur}{m}{n}
      \define@mathgroup\mv@bold\upmath@group{eur}{b}{n}
      \edef\UPM{\hexnumber\upmath@group}
      \new@mathgroup\amsa@group
      \define@mathgroup\mv@normal\amsa@group{msa}{m}{n}
      \define@mathgroup\mv@bold\amsa@group{msa}{m}{n}
      \edef\AMSa{\hexnumber\amsa@group}
      \makeatother
      \mathchardef\upi="0\UPM19
      \mathchardef\umu="0\UPM16
      \mathchardef\upartial="0\UPM40
      \mathchardef\leqslant="3\AMSa36
      \mathchardef\geqslant="3\AMSa3E

       \let\le=\leqslant

    \fi
  \fi
\fi 

\ifnfsstwo
  \DeclareMathAlphabet{\mathbfit}{OT1}{cmr}{bx}{it}
  \SetMathAlphabet\mathbfit{bold}{OT1}{cmr}{bx}{it}
  \DeclareMathAlphabet{\mathbfss}{OT1}{cmss}{bx}{n}
  \SetMathAlphabet\mathbfss{bold}{OT1}{cmss}{bx}{n}
  \ifAMStwofonts
    \ifCUPmtlplainloaded \else
      \DeclareSymbolFont{UPM}{U}{eur}{m}{n}
      \SetSymbolFont{UPM}{bold}{U}{eur}{b}{n}
      \DeclareSymbolFont{AMSa}{U}{msa}{m}{n}
      \DeclareMathSymbol{\upi}{0}{UPM}{"19}
      \DeclareMathSymbol{\umu}{0}{UPM}{"16}
      \DeclareMathSymbol{\upartial}{0}{UPM}{"40}
      \DeclareMathSymbol{\leqslant}{3}{AMSa}{"36}
      \DeclareMathSymbol{\geqslant}{3}{AMSa}{"3E}

       \let\le=\leqslant

    \fi
  \fi
\fi 

\ifCUPmtlplainloaded \else
  \ifAMStwofonts \else 
    \def\upi{\pi}
    \def\umu{\mu}
    \def\upartial{\partial}
  \fi
\fi

\title{Goodness-of-Fit Tests to study the Gaussianity 
of the MAXIMA data}
\author[] {L. Cay\'on$^{1,2}$
, F. Arg\"ueso$^{3}$
, E. Mart\'\i nez-Gonz\'alez$^{1}$ and J.L. Sanz$^{1}$.\\
1. Instituto de F\'\i sica de Cantabria, Fac. Ciencias, Av. los
	Castros s/n, 39005 Santander, Spain\\
2. Physics Department, Purdue University, 525 Northwestern Avenue, West Lafayette, IN 47907-2036, USA\\
3. Dpto. de Matem\'aticas, Universidad de Oviedo, c/ Calvo Sotelo s/n, 33007 Oviedo, Spain\\
}

\date{\today}

\pagerange{\pageref{firstpage}--\pageref{lastpage}}
\pubyear{2001}

\begin{document}

\maketitle

\label{firstpage}

\begin{abstract}
Goodness-of-Fit tests, including Smooth ones, 
are introduced and applied to detect non-Gaussianity in Cosmic Microwave Background simulations. We study the power of three different tests: the 
Shapiro-Francia test (1972), the uncategorised smooth test developed by Rayner and
Best(1990) and the Neyman's Smooth Goodness-of-fit test for composite
hypotheses (Thomas $\&$ Pierce 1979). The Smooth Goodness-of-Fit tests are
designed to be sensitive to the presence of ``smooth'' deviations from
a given distribution. We study the power of these tests based on the 
discrimination between Gaussian and non-Gaussian simulations.
Non-Gaussian cases are simulated using the Edgeworth expansion and assuming pixel-to-pixel independence. 
Results show these tests behave similarly and are more powerful than tests directly
based on cumulants of order 3, 4, 5 and 6. 
We have applied these tests to the released MAXIMA data.
The applied tests are built to be powerful against 
detecting deviations from univariate Gaussianity. The 
Cholesky matrix corresponding to signal (based on an assumed 
cosmological model) plus noise is used to
decorrelate the observations previous to the analysis. 
Results indicate that the 
MAXIMA data are compatible with Gaussianity.

\end{abstract}



\section{Introduction}

The detection of non Gaussianity in Cosmic Microwave Background (CMB) maps
will question the validity of Standard Inflationary theories. These theories
assume the existence of a single scalar field as well as linear theory, 
to generate the cosmological perturbations that will later develop into
the structures observed in the Universe. Some alternative scenarios 
will include the presence of topological defects (Durrer 1999) or isocurvature 
fluctuations (Peebles 1999a,b), multi-field inflation models (Bernardeau \& Uzan 2002 and references therein) and 
stochastic inflationary 
scenarios generating features in the inflaton potential (Starobinsky 1986). Moreover, in a recent work by Acquaviva et al. (2002) it is shown
how the inclusion of secondary effects will modify the predictions of 
one single scalar field theories.

All the above alternatives to Standard Inflation will result in non Gaussian
 CMB temperature fluctuations. The type and amount of non Gaussianity 
to be observed in CMB maps is under study at the moment. There have been 
several works exploring the implications on CMB observations 
of different physical mechanisms 
that will generate non Gaussianity (Komatsu \& Spergel 2001, Landriau \& Shellard 2002, Acquaviva et al. 2002, Gupta et al. 2002, Gangui et al. 1994).
Tests of a scenario including a quadratic term in the gravitational 
potential have been performed on COBE-DMR CMB data (Komatsu et al. 2002,
Cay\'on et al. 2002). The poor resolution of these data does not provide 
a very good constraint of the non-linear coupling parameter accounting for 
the contribution of the quadratic term. However, it can be concluded that
the method based on the Spherical Mexican Hat wavelet provides a 
better constraint than the one based on the bispectrum.

At present there is a large effort to implement different statistical tools
that will allow us in the future to test the Gaussianity of observed
CMB data. The power of the different methods will vary depending on the 
type of non Gaussianity present in the data. In this paper we propose 
three Goodness-of-fit tests and study their power on simulations. We simulate
Gaussian and non Gaussian maps. The later are performed using the 
Edgeworth expansion. This expansion was for the first time used to simulate
non Gaussian CMB maps by Mart\'\i nez-Gonz\'alez et al. 2002.
The proposed methods are specially well suited to analyse data covering a 
region of the sky and not necessarily taken on a regular grid.
We have applied them to the recently released MAXIMA
data (Balbi et al. 2000, Hanany et al. 2000). The MAXIMA data have already been
tested against Gaussianity by Wu et al. 2001 and Santos et al. 2002a,b. Both 
works conclude that the data are compatible with Gaussianity.

The paper is organized as follows. Section 1 is dedicated to present the 
Goodness-of-fit tests as well as to test them on non Gaussian simulations 
generated using the Edgeworth expansion. An application to the MAXIMA data 
is presented in Section 2. Discussion and conclusions are included in Section 3.

\section{Goodness-of-fit Statistics}

\begin{figure}
\epsfxsize=84mm
\epsffile{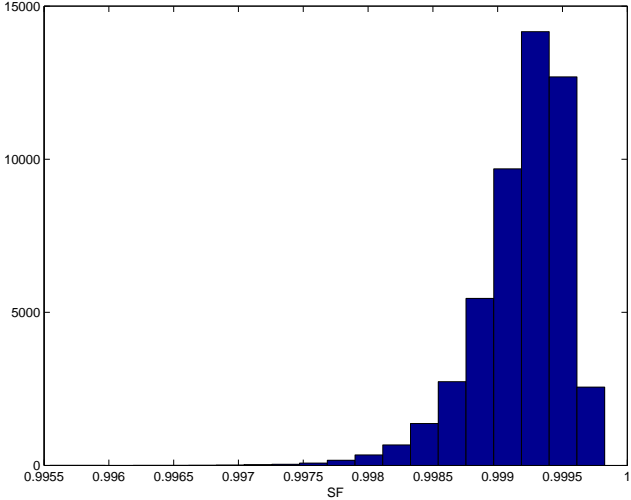}
 \caption{Distribution the Shapiro-Francia statistic 
obtained from 50000 independent Gaussian simulations of 
maps with 2164 pixels (independent pixel-to-pixel).
}
 \label{f1}
\end{figure}

Given a sample of uncorrelated and normalized (zero mean, dispersion one)
CMB data, 
the question we want to answer is 
``how well the data agree with the population of a Gaussian 
distribution $N(0,1)$'' (the methods here used are also suited for testing
a composite hypothesis where mean and dispersion are not specified).

Many goodness-of-fit methods have been 
developed to test normality (for a review see D'Agostino $\&$ 
Stephens 1986). Out of all we have chosen to apply the 
Shapiro-Francia one (Shapiro \& Francia 1972), a modification of the Shapiro-Wilk test for 
large data sets. Implementation of the Shapiro-Francia requires
the following steps:\hfill\break
\noindent 1) - Estimation of the 1 dimensional array $\vec c$ corresponding 
to the expected sorted values obtained 
from independent Gaussian simulations $N(0,1)$. We define $\vec b=\vec c/\vert\vert c\vert\vert^{1/2}$ \hfill\break
\noindent 2) - For a given sorted data set $\vec x$ , the Shapiro-Francia statistic $SF$ is defined as follows
$$SF={{(\sum_{i=1}^{N}b_ix_i)^2}\over{N\sigma^2}},$$
\noindent where $N$ is the number of data and $\sigma$ is the 
dispersion. The expected value of this statistic for the Gaussian 
distribution is very close to one. Deviations from Gaussianity will result in
values smaller than one. The distribution of the SF statistic for
independent Gaussian realizations is presented in Figure 1.

Smooth goodness-of-fit methods have been constructed as powerful tests against
distributions that might deviate ``smoothly'' from the normal $N(0,1)$
one. These methods are sensitive to the presence of skewness ($S$), kurtosis ($K$) and
higher order moments. 
Some of them are defined based on
orthonormal functions (Rayner $\&$ Best 1990) whereas others 
make use of powers of the distribution function (Thomas $\&$ Pierce 1979). 
The uncategorised smooth models proposed by Rayner $\&$ Best (1990) $S_k$ make use of the 
Hermite-Chebyshev polynomials $P_n$ and are defined by
$$S_k=\sum_{i=1}^k (\sum_{j=1}^Nh_i(x_j)/\sqrt(N))^2,$$
\noindent where $h_i(x_j)=P_i(x_j)/\sqrt(n\!)$. 
The statistic $S_k$ is related to cumulants of
order $\le k$, such that for example:
$$S_1=N<x>^2, ~~~S_2=S_1+(N/2)(<x^2>-1)^2,$$
$$S_3=S_2+(N/6)(<x^3>-3<x>)^2,$$
$$S_4=S_3+(N/24)(<x^4>-6<x^2>+3)^2,$$
\noindent where $<>$ denotes the average. And if 
$<x>=0$ and $<x^2>=1$, then
$S_3=(N/6)S^2$ and $S_4=S_3+(N/24)K^2$.
The $S_k$ statistic is 
distributed as a $\chi_k^2$ for the Gaussian case. 
We also make use in this paper of the smooth goodness-of-fit test
$W_k$ proposed by Thomas $\&$ Pierce (1979) as a modification of Neyman's one.
This test is built on powers of the normal distribution function and compares
sample means of these quantities with the expected values under the 
null hypothesis that the data correspond to a population sample
of the normal distribution. 
$$W_k = \sum_{i = 1}^{k} \bigl[\sum_{j = 1}^{i} a_{ij}u_j\bigr]^2 , k = 1, 2, 3, 4,...$$
$$u_j\equiv \frac{1}{N^{1/2}}{\sum}_{r = 1}^{N}  (y^j(x_r) - \frac{1}{1 + j})$$
where for the normal distribution $y(x_r)\equiv erf(\frac{x_r - \mu}{\sigma})$, being  
$\mu ,\sigma $ the mean and dispersion, and the coefficients $a_{ij}$ are given in Table 3  by 
Thomas \& Pierce (1979) (e. g. $a_{11} = 16.3172, a_{21} = - a_{22} = -27.3809$). Therefore, 
for $k = 1, 2$, for example, the statistics are
$$W_1 = \frac{16.3172^2}{N}\bigl[{\sum}_{r = 1}^{N} (y(x_r) - \frac{1}{2})\bigr]^2,$$
$$W_2 = W_1 + \frac{27.3809^2}{N}\bigl[{\sum}_{r = 1}^{N} (y^2(x_r) - \frac{1}{3} - y(x_r) + \frac{1}{2})\bigr]^2.$$
These statistics, under the null hypothesis, are distributed as $\chi^2_k$.
The distributions of values of the two smooth goodness-of-fit statistics for 
50000 independent Gaussian
realizations are presented in Figures 2 and 3. In both cases, and in the 
examples and applications presented below, the data are renormalized to 
zero mean and unit variance before the tests are applied. 
Because of that the $S_k$ statistics appear as $\chi_{k-2}^2$ distributions.
The distributions of the two statistics converge to the expected ones. 
However one can see that the $W_k$ statistic converges faster to the
expected distribution than the $S_k$.

\begin{figure*}
 \epsffile{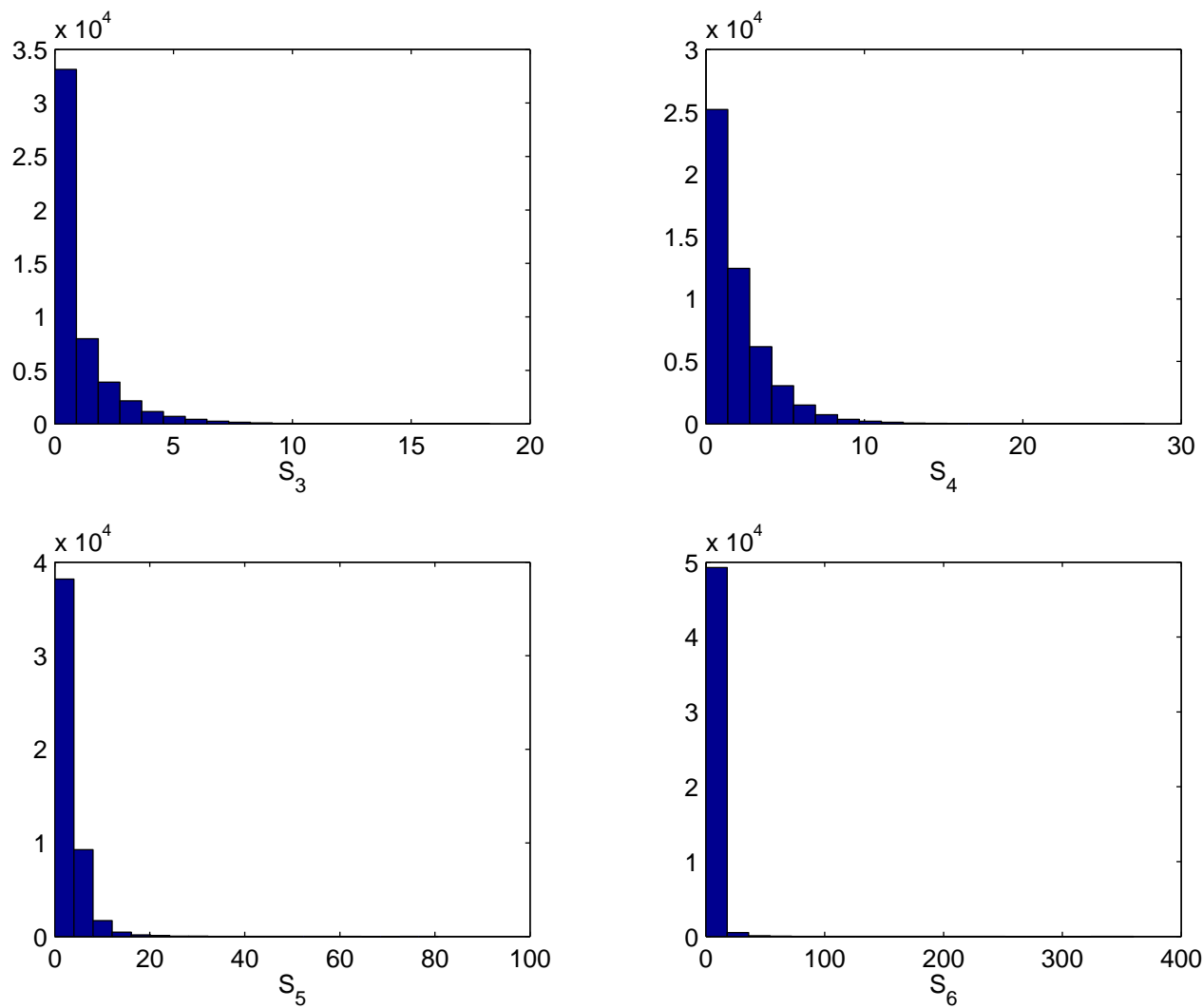}
 \caption{Distribution of $S_k$ smooth-goodness-of-fit statistics.
From left to right, top to bottom $S_k$  corresponding to $S_3, S_4, S_5, S_6$.
Distributions obtained from 50000 independent Gaussian simulations of 
maps with 2164 pixels (pixel-to-pixel independent).
}
 \label{f1}
\end{figure*}

\begin{figure*}
 \epsffile{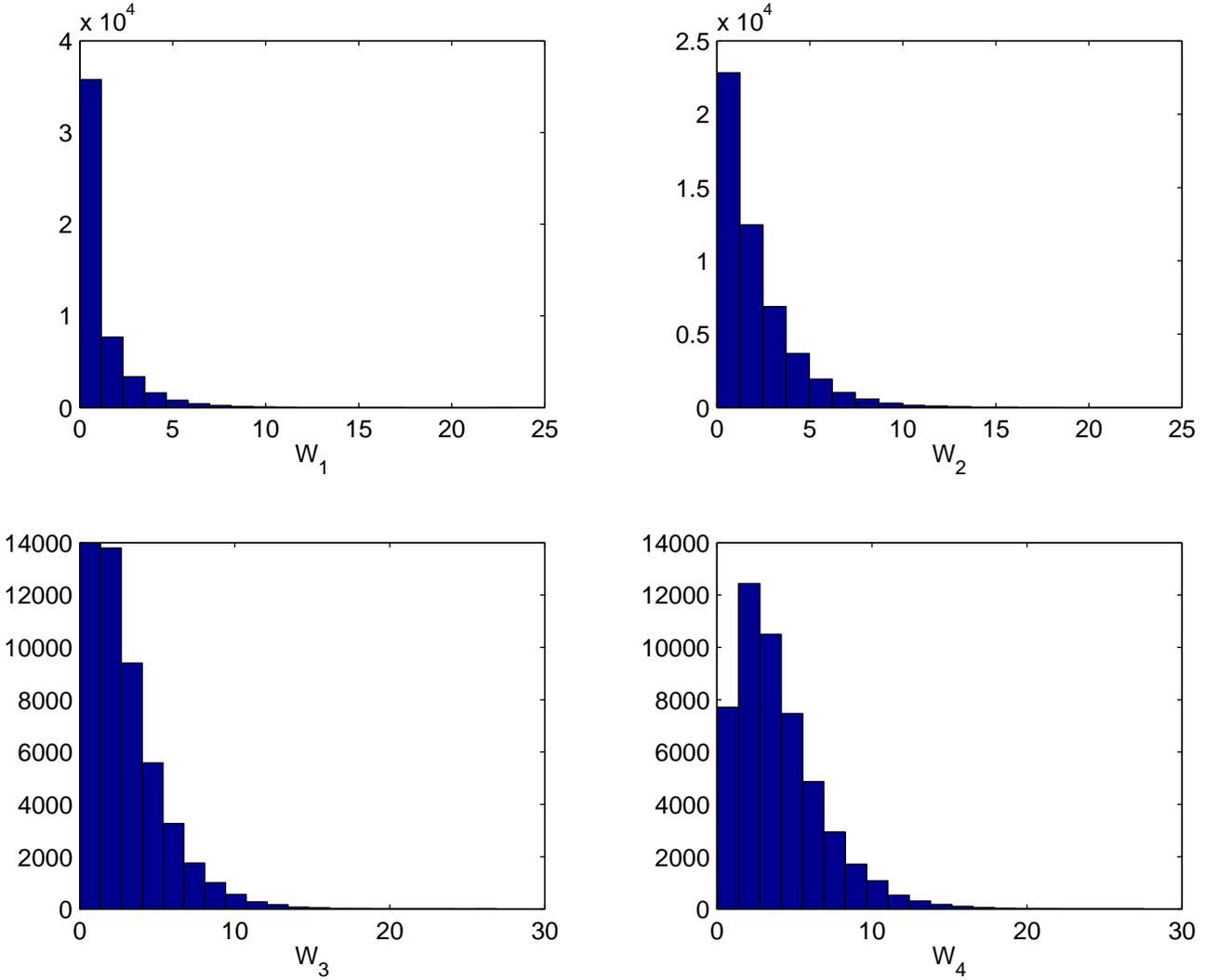}
 \caption{Distribution of $W_k$ smooth-goodness-of-fit statistics.
From left to right, top to bottom $W_k$ corresponding to $W_1, W_2, W_3, W_4$.
Distributions obtained from 50000 independent Gaussian simulations of 
maps with 2164 pixels (pixel-to-pixel independent).
}
 \label{f1}
\end{figure*}

As an example of how well these methods work on separating Gaussian
from nonGaussian data, we have applied them to simulated non Gaussian
data following distributions as in Mart\'\i nez-Gonz\'alez et al. (2002).
An array with 2164 independent pixels constitutes a simulation. 
The chosen number of pixels fits the number of central
pixels in the MAXIMA map, later selected for analysis. 
The pixel value is drawn from a non Gaussian distribution obtained 
through the Edgeworth expansion characterised by a given skewness and kurtosis.
Given an input skewness and kurtosis, the mean and dispersion value for
these two statistics in the resulting simulations are given in Table 1,
as well as the skewness and kurtosis power to discriminate between
Gaussian and non Gaussian distributions.  
10000 simulations were performed and the power of the Smooth Goodness 
of fit statistics presented in this paper 
is given for several input skewness and kurtosis values in Table 2 and 
Table 3. 
As one can see from these results, most of the presented goodness-of-fit 
statistics have more power than the directly calculated cumulants.
The $W_2$ statistic is the one with higher discriminating power
in most of the cases. 
Even so, the result is very dependent on the underlying 
distribution.

\begin{table*}
  \begin{minipage}{170mm}
  \begin{center}
  \caption[]{Average and dispersion for the skewness and 
kurtosis values obtained from 10000 simulations drawn from 
Edgeworth expansions assuming skewness and kurtosis
values denoted by $S\&K(input)$. The power of these two statistics is also
given in columns 4 and 5.}
  \label{tab1}
  \begin{tabular}{c|c|c|c|c}\hline
  S$\&$K(input)& Mean/Disp (S)& Mean/Disp (K) & Power(95/99$\%$) (S)& Power(95/99$\%$) (K)\cr
  \hline\hline 
	0.0$\&$0.4&5.95e-4/0.0607&0.3170/0.1205&7.86/2.13&87.85/65.57\cr
	0.1$\&$0.0&0.0965/0.0503&-0.0342/0.0938&57.51/28.36&1.65/0.19\cr
	0.1$\&$0.3&0.0982/0.0581&0.2309/0.1157&57.49/32.25&67.26/37.05\cr
	0.1$\&$0.4&0.0968/0.0604&0.3180/0.1239&56.22/31.6&87.72/65.19\cr
	0.1$\&$0.5&0.0976/0.0624&0.4061/0.1273&56.46/32.67&97.08/86.75\cr
	0.1$\&$0.7&0.0976/0.0624&0.5820/0.1391&57.27/35.62&99.96/99.17\cr
	0.3$\&$0.3&0.2944/0.0565&0.2323/0.1320&99.96/99.86&64.90/38.87\cr

  \hline\hline
  \end{tabular}
  \end{center}
\end{minipage}
\end{table*}

\begin{table*}
  \begin{minipage}{170mm}
  \begin{center}
  \caption[]{Power at $95\%$ and $99\%$ confidence level for the different
statistics presented in this work (notation used for the table 
``$95\%/99\%$''). These results are
based on 10000 Gaussian and non Gaussian simulations.
The non Gaussian ones were obtained from the Edgeworth expansion 
for different values of skewness and kurtosis (skewness $S$ and kurtosis $K$ input values
indicated in column 1).}
  \label{tab1}
  \begin{tabular}{c|c|c|c|c|c|c|c|c}\hline
  S$\&$K& $S_3$& $S_4$&$S_5$&$S_6$&$W_1$&$W_2$&$W_3$&$W_4$\cr
  \hline\hline 
	0.0$\&$0.4&8.95/2.46&74.04/53.18&66.48/36.26&71.98/23.32&6.80/1.67&83.03/64.75&77.20/58.76&74.38/52.97\cr
	0.1$\&$0.0&44.62/20.51&33.4/12.93&23.63/5.11&16.7/0.65&41.13/20.26&33.22/14.31&29.16/12.38&25.61/9.11\cr
	0.1$\&$0.3&46.52/25.03&68.46/47.10&62.03/31.30&61.27/15.97&45.99/25.31&74.22/53.98&68.43/48.04&65.38/42.2\cr
	0.1$\&$0.4&45.58/24.99&84.35/67.6&79.38/52.71&81.91/35.99&46.35/25.42&90.79/77.35&87.24/72.32&84.81/66.26\cr
	0.1$\&$0.5&46.35/26.03&94.56/86.13&92.80/74.72&95.00/62.99&47.64/27.23&98.27/93.92&97.04/91.54&96.23/88.04\cr
	0.1$\&$0.7&47.96/28.82&99.75/98.92&99.56/97.05&99.91/97.06&50.88/31.07&99.99/99.95&99.97/99.89&99.96/99.75\cr
	0.3$\&$0.3&99.95/99.75&99.90/99.51&99.86/99.58&99.80/92.81&99.96/99.85&99.95/99.72&99.92/99.46&99.88/99.00\cr
  \hline\hline
  \end{tabular}
  \end{center}
\end{minipage}
\end{table*}

\begin{table*}
  \begin{minipage}{170mm}
  \begin{center}
  \caption[]{Power at $95\%$ and $99\%$ confidence level for the 
Shapiro-Francia statistic (notation used for the table 
``$95\%/99\%$''). These results are
based on 10000 Gaussian and non Gaussian simulations.
The non Gaussian ones were obtained from the Edgeworth expansion 
for different values of skewness and kurtosis (skewness $S$ and kurtosis $K$ input values
indicated in column 1).}
  \label{tab1}
  \begin{tabular}{c|c}\hline
  S$\&$K&SF\cr
  \hline\hline 
	0.0$\&$0.4&70.82/47.30\cr
	0.1$\&$0.0&30.97/13.69\cr
	0.1$\&$0.3&65.21/42.45\cr
	0.1$\&$0.4&83.37/64.62\cr
	0.1$\&$0.5&95.35/86.27\cr
	0.1$\&$0.7&99.90/99.56\cr
	0.3$\&$0.3&99.94/99.66\cr
  \hline\hline
  \end{tabular}
  \end{center}
\end{minipage}
\end{table*}

The cases simulated above are based on an Edgeworth expansion 
considering all cumulants of order greater than 2 equal to zero, except for 
those of order 3 and 4. The smooth goodness of fit statistics combine 
information from cumulants of any order and in particular of orders
below 7. We have performed simulations based on the Edgeworth 
expansion including non null cumulants up to order 6. The simulated maps 
do not preserve the input cumulant values. Mean and dispersion values
of the cumulants calculated out of 10000 simulations 
are presented in Table 4. The power of the cumulants to differentiate
between a Gaussian and another distribution (based on the Edgeworth
expansion)  is given at the $95\%$ in Table 5.

\begin{table*}
  \begin{minipage}{170mm}
  \begin{center}
  \caption[]{Average and dispersion for the skewness,
kurtosis, 5th and 6th order cumulants obtained from 10000 simulations drawn from 
Edgeworth expansions assuming skewness, kurtosis, 5th and 6th order
cumulant
values denoted by $S\&K\&k5\&k6(input)$.}
  \label{tab1}
  \begin{tabular}{c|c|c|c|c}\hline
  S$\&$K$\&$k5$\&$k6 (input)& Mean/Disp (S)& Mean/Disp (K) & 
Mean/Disp (k5)& Mean/Disp (k6)\cr
  \hline\hline 
	0.1$\&$0.4$\&$1.0$\&$3.0&0.0955/0.0673&0.3169/0.1540&0.8156/0.3246&1.3604/0.67157\cr
	0.1$\&$0.3$\&$1.0$\&$3.0&0.1264/0.0617&0.1475/0.1421&1.0634/0.2610&1.2203/0.5402\cr
	0.1$\&$0.4$\&$0.8$\&$3.0&0.0937/0.0677&0.3160/0.1563&0.6076/0.3450&1.4349/0.6925\cr
	0.1$\&$0.3$\&$0.8$\&$3.0&0.1397/0.0602&0.1123/0.1417&0.9523/0.2714&1.2030/0.5468\cr

  \hline\hline
  \end{tabular}
  \end{center}
\end{minipage}
\end{table*}

\begin{table*}
  \begin{minipage}{170mm}
  \begin{center}
  \caption[]{Power of the skewness, kurtosis, 5th and 6th order cumulants 
at the $95\%$ confidence level.
Results drawn from 10000 simulations based on 
Edgeworth expansions assuming skewness, kurtosis, 5th and 6th order
cumulant
values denoted by $S\&K\&k5\&k6(input)$.}
  \label{tab1}
  \begin{tabular}{c|c|c|c|c}\hline
  S$\&$K$\&$k5$\&$k6 (input)& Power (S)& Power (K) & 
Power (k5)& Power (k6)\cr
  \hline\hline 
	0.1$\&$0.4$\&$1.0$\&$3.0&54.91&81.48&91.09&73.10\cr
	0.1$\&$0.3$\&$1.0$\&$3.0&73.52&41.46&99.72&69.97\cr
	0.1$\&$0.4$\&$0.8$\&$3.0&53.94&81.27&75.17&76.32\cr
	0.1$\&$0.3$\&$0.8$\&$3.0&80.72&32.17&98.86&68.17\cr

  \hline\hline
  \end{tabular}
  \end{center}
\end{minipage}
\end{table*}

The smooth goodness of fit statistics as well as the Shapiro-Francia test
have been calculated for the simulated maps including cumulants
up to order 6. Powers at the $95\%$ confidence level are presented 
in Table 6. As one can see in all cases the Shapiro-Francia test
is the most powerful of all the implemented tests. It is difficult to 
asses whether some tests will be more convenient than others when trying 
to detect non Gaussianity. Moreover, as pointed out by Bromley \&
Tegmark (1999) even if non Gaussianity is detected by one statistical 
method, the confidence level has to be established taking into 
account all the methods applied.

\begin{table*}
  \begin{minipage}{170mm}
  \begin{center}
  \caption[]{Power at $95\%$ confidence level for the different
statistics presented in this work. These results are
based on 10000 Gaussian and non Gaussian simulations.
The non Gaussian ones were obtained from the Edgeworth expansion 
for different values of skewness, kurtosis, 5th and 6th order cumulants
(input values for skew $S$, kurt $K$, 5th $k5$ and 6th $k6$ order cumulants 
indicated in column 1.}
  \label{tab1}
  \begin{tabular}{c|c|c|c|c|c|c|c|c|c}\hline
  S$\&$K$\&$k5$\&$k6 & $S_3$ & $S_4$ & $S_5$ & $S_6$ & $W_1$ & $W_2$ & $W_3$ & $W_4$ & SF\cr
  \hline\hline 
	0.1$\&$0.4$\&$1.0$\&$3.0&45.92&78.11&93.63&97.54&9.61&17.04&37.68&61.16&99.91\cr
	0.1$\&$0.3$\&$1.0$\&$3.0&64.77&60.28&98.87&98.98&13.30&14.18&37.29&41.48&99.98\cr
	0.1$\&$0.4$\&$0.8$\&$3.0&44.85&78.46&87.77&94.78&12.96&20.30&35.95&52.81&99.36\cr
	0.1$\&$0.3$\&$0.8$\&$3.0&72.82&64.30&97.58&97.54&22.13&21.61&46.99&44.46&99.88\cr

  \hline\hline
  \end{tabular}
  \end{center}
\end{minipage}
\end{table*}

\section{MAXIMA data analysis. Results}

The goodness-of-fit tests previously described 
are optimal for testing univariate Gaussian distributions. We therefore 
first of all transform the MAXIMA data by multiplying it by the 
inverse of the Cholesky matrix corresponding to signal plus noise.
We first calculate the
Cholesky decomposition 
corresponding to the ``data correlation matrix''. To do that
a cosmological model fitting the data has been assumed. 
MAXIMA data is better fitted to a cosmological model characterised by
$\omega_b=0.105$, $\omega_c=0.595$, $\omega_{\Lambda}=0.3$ and
$h=0.53$ (Balbi et al. 2000). Simulations 
of this model are used to calculate the signal correlation matrix.
The so called ``data correlation matrix'' is the sum of the signal and noise
correlation matrices. 

Once the data respond to independent values drawn from a normal 
distribution with mean zero and dispersion one $N(0,1)$, 
we calculate 
the above introduced smooth goodness-of-fit statistics as well as the 
Shapiro-Francia one. 
Noise levels are specially high at some pixels resulting in large
correlation values off the diagonal. Most of these
pixels appear to be in the border of the observed region.
To make sure the previous test
is not dominated by noise we have selected pixels in the central area 
of the one covered by MAXIMA observations. We select pixels with 
right ascention in the range $226.47-238.24$ degs and
declination ranging from $55.567$ degs to $61.7$ degs. These pixels amount
to a total of 2164.
In order to see if the MAXIMA data are compatible with
Gaussianity we have simulated 50000 maps with 2164 pixels independently 
generated from a $N(0,1)$ distribution. The statistic values 
obtained from the data as well as the 
probability of having Gaussian values larger than the MAXIMA ones are
presented in Table 7.
One should keep in mind that in the case of
the Shapiro-Francia test, this is a confidence level taken from the left.
As one can see, the MAXIMA data are compatible with Gaussianity.

\begin{table*}
  \begin{minipage}{170mm}
  \begin{center}
  \caption[]{Statistic values obtained for the MAXIMA data and probability of being drawn from a Gaussian distribution (Prob $>= MAXIMA$) from independent Gaussian simulations}
  \label{tab1}
  \begin{tabular}{c|c|c|c|c}\hline
  statistic& data value & Prob$\%$\cr
  \hline\hline 
	S&0.035&25.23\cr
	K&0.058&28.04\cr
	k5&-0.276&89.43\cr
	k6&-0.068&48.32\cr
	$S_3$&0.446&50.29\cr
	$S_4$&0.717&69.74\cr
	$S_5$&2.091&52.34\cr
	$S_6$&2.101&64.37\cr
	$W_1$&0.948&33.23\cr
	$W_2$&1.050&59.64\cr
	$W_3$&1.170&76.29\cr
	$W_4$&1.177&88.40\cr
	SF&0.9994&30.06\cr
  \hline\hline
  \end{tabular}
  \end{center}
\end{minipage}
\end{table*}

The process we follow to decorrelate the observations could 
be thought to introduce some artifacts that could make the comparison
with independent $N(0,1)$ simulations not appropiate. 
To answer this question we simulate data taking into account
MAXIMA observational constraints as well as the noise information.
Afterwards these simulations are decorrelated following the same steps
as in the the real data case. The statistical tests are applied on these
decorrelated simulations and compared with the results obtained
from the decorrelated MAXIMA data. 

As a first step we ``tried'' simulating CMB skies as those seen by
MAXIMA. Those simulations included signal plus noise.
The quoted word refers to the signal simulations. To simplify the 
simulation process we have based the simulations on the HEALpix 
pixelization (Gorski, Hivon \& Wandelt 1999). This does not exactly reproduce the observational
grid but we consider the approach good enough for the proposed test.
These maps are simulated following these steps:\hfill\break
\noindent 1) The $a_{lm}$ coefficients are generated assuming the 
power spectrum corresponding to the cosmological model that
best fit the MAXIMA data (Balbi et al. 2000) multiplied by the 
beam pattern.\hfill\break
\noindent 2) We use the HEALpix subroutines to generate a map with
the obtained $a_{lm}$s. The maximum $l$ corresponds to a $nside=512$,
that is, the generated pixels are $7\times 7$ arcmin$^2$. MAXIMA pixels
are $8\times 8$ arcmin$^2$. Moreover, the MAXIMA pixelization does not
agree with the HEALpix pixelization. The simulated pixel
values are assigned to MAXIMA pixels based on their right 
ascention and declination. Since the processing time is 
quite long we only performed 300 simulations.
Noise simulations are obtained by multiplying  
a pixel-to-pixel independently generated (from a Gaussian distribution
$N(0,1)$) map by the Cholesky matrix corresponding to the noise 
correlation matrix. 

The simulated maps are afterwards multiplied by the inverse of the 
Cholesky matrix as it is done with the data.
Results for the different statistics evaluated 
on the MAXIMA data and probability of these values to be drawn from
a Gaussian distribution are presented in Table 8. The MAXIMA values
are in agreement with Gaussian ones as was obtained in the previous case. 
The probability values obtained for the different statistics in Tables 7 and 8
are of the same order. One can however notice a larger difference in the
case of the $K$ statistic. We have checked the $K$ values obtained in
both cases (for Tables 7 and 8). We have done the exercise of obtaining
the distribution of $\vert K \vert$ values directly from 
the $K$ ones (case 1) and combining the $S_3$ and $S_4$ 
values as indicated in section 2 (case 2). In case 1, the 
MAXIMA value is $\vert K_{MAXIMA}\vert =0.058$ (as indicated in Table 7)
and the probability of getting values greater or equal than this one
is $58.15\%$ for simulations in Table 7 and $65.00\%$ for simulations in
Table 8. In case 2, the MAXIMA value is $\vert K_{MAXIMA}\vert =0.055$ and
the probability of getting values greater or equal than this one
is $60.18\%$ for simulations in Table 7 and $65.00\%$ for simulations in
Table 8. There is therefore compatibility between these two cases.
Moreover, the distribution of the absolute value of the $K$ statistic 
seems not to change much between simulations in Table 7 and those
in Table 8. 
What might be happening is that the distribution of $K$ values converges 
slowly and therefore 300 simulations, in the case of those in Table 8, might
not be enough to asses a precise probability value. Nevertheless this result
does not affect our final conclusion, that the MAXIMA data are found
to be compatible with Gaussianity under the Goodness-of-fit tests
applied in this work.

Finally, we would like to note the importance of decorrelating the
data in order to asses its departure from Gaussianity, by looking at
the Goodness-of-fit statistics introduced in this paper. As already
mentioned, these are statistics designed to test univariate Gaussianity.
Applying these methods to correlated data will only have partial meaning
and their power will be diluted.
We should however mention the fact that the decorrelation procedure
will modify the distribution of the data we are analysing. It is
difficult to quantify this effect as it would depend on the deviations
from Gaussianity present in the analysed data as well as on the
corresponding Cholesky matrix. Just as an example we have applied all
the statistical tests discussed in this work to two cases in which 
a certain amount of non-Gaussianity is introduced in correlated
simulations. The results are presented in Table 9. The power of the statistics 
in distinguishing Gaussian from non-Gaussian simulations is shown in
columns 3 and 4 for the two cases considered. In each column, the first
number represents the power at the $95\%$ c.l. when looking at correlated simulations. The second
number indicates the power at the $95\%$ c.l. after decorrelating the simulations (the inverse of the Cholesky matrix corresponding to
MAXIMA data is used for this). The Gaussian simulations are done as explained in the fourth paragraph
of this Section, following MAXIMA's constraints. Each of the non-Gaussian ones
consists of the sum of a Gaussian simulation plus a non-Gaussian one 
with the same correlation (the fact that we have a sum of two
simulations with the same correlation is taken into account when
decorrelating). The non-Gaussian simulations are generated 
based on an Edgeworth expansion as the ones in Section 2, afterwards multiplied
by the Cholesky matrix corresponding to the MAXIMA data.
As can be seeing from the Table, any of the suggested methods have a 
larger power after decorrelating, even if then distributions have been
modified in the process. 

%
\begin{table*}
  \begin{minipage}{170mm}
  \begin{center}
  \caption[]{Probability of the MAXIMA data of being drawn from a Gaussian distribution (
Prob $>= MAXIMA$). Simulations consist of the sum of signal plus noise following the 
MAXIMA observational constraints. Both the data and the simulations are
decorrelated (as explained in the text) before they are analysed.}
  \label{tab1}
  \begin{tabular}{c|c|c}\hline
  statistic& Prob$\%$\cr
  \hline\hline 
	S&27.67\cr
	K&48.33\cr
	k5&84.33\cr
	k6&46.33\cr
	$S_3$&51.33\cr
	$S_4$&73.00\cr
	$S_5$&57.67\cr
	$S_6$&68.33\cr
	$W_1$&34.00\cr
	$W_2$&63.67\cr
	$W_3$&76.67\cr
	$W_4$&89.00\cr
	SF&27.00\cr
  \hline\hline
  \end{tabular}
  \end{center}
\end{minipage}
\end{table*}

\begin{table*}
  \begin{minipage}{170mm}
  \begin{center}
  \caption[]{Power of the different statistical tests to distinguish
Gaussian from non-Gaussian simulations before and after
decorrelating (bef/aft). For the two cases considered, the non-Gaussian simulations are generated
as explained in the text and based on Edgeworth expansions
with $S=0.5,K=0.5$ (column 2) and $S=0.7,K=0.7$ (column 3).}
  \label{tab1}
  \begin{tabular}{c|c|c|c}\hline
  statistic& Power($95\%$) bef/aft $S=0.5,K=0.5$& Power($95\%$) bef/aft $S=0.7,K=0.7$ \cr
  \hline\hline 
	S&20.66/86.33&19.66/99.00\cr
	K&26.00/6.66&23.33/11.66\cr
	k5&18.33/8.66&14.33/8.33\cr
	k6&25.33/6.00&22.66/4.66\cr
	$S_3$&19.99/76.00&23.00/96.00\cr
	$S_4$&28.66/53.00&24.66/84.67\cr
	$S_5$&26.66/28.33&23.00/62.00\cr
	$S_6$&26.00/18.33&23.66/41.33\cr
	$W_1$&13.66/78.00&14.99/98.33\cr
	$W_2$&18.00/66.33&19.66/93.33\cr
	$W_3$&17.66/69.00&19.66/92.66\cr
	$W_4$&22.66/63.00&21.66/91.33\cr
	SF&24.00/49.00&20.33/85.66\cr
  \hline\hline
  \end{tabular}
  \end{center}
\end{minipage}
\end{table*}

\section{Discussion and Conclusions}

The future detection of non Gaussianity in CMB data will rely on 
the power of the applied statistical methods. At present, it is not
an easy task to establish which methods will be more powerful.
Physically motivated deviations from Gaussianity generated in 
theories alternative to the Standard Inflationary one are still not well
characterised. 
It is however needed to know which methods could be better suited to
detect certain types of non Gaussianity even if they are tested on
toy model simulations.

We have implemented goodness-of-fit statistics developed to detect deviations from a given distribution, in our case from the Gaussian one. Three different methods have been tested against simulations including deviations from
the Gaussian distribution. The selected methods were developed 
by Shapiro \& Francia (1972), Thomas \& Pierce (1979) and Rayner \& Best (1990). The last two ones belong to the so called smooth goodness-of-fit tests. 
The performance of these methods was checked on simulations based on the
Edgeworth expansion including distortions produced by the presence of 
cumulants of order higher than two. A strong conclusion can not be drawn from
this exercise. If only
skewness and kurtosis are present, the statistic $W_2$ developed by Thomas \& Pierce (1979) has more power than the rest of applied statistics.  
The presence of cumulants of order 5 and 6 seems to be better detected by the Shapiro-Francia test. 

Several statistical methods have already been applied to the MAXIMA data in the
search for non Gaussianity. Wu et al. (2001) calculated moments, cumulants, Minkowski functionals, Kolmogorov and $\chi^2$ tests of the real and
Wiener filtered data as well as of the eigenmodes and signal-whitened data.
Santos et al. (2002) obtained the bispectrum value for these data. In both cases comparison with Gaussian predictions confirmed the compatibility of the
MAXIMA data with Gaussianity. We have added three more tests to the ones applied
to MAXIMA (also see Aliaga et al. (2003) where constraints on $S$ snd $K$
are impossed based on that data). The tests are optimal in the case of non-correlated data.
Moreover, they can be applied even in cases in which the observations
are not taken on a regular grid. 
As a first step in our method we have decorrelated the observations by 
multiplying by the inverse of the corresponding Cholesky matrix. This is
feasible in this case in which only a region of the whole sky was covered 
as any matrix operation is computationally very expensive. 
Future application of
these methods could in any case be done by decorrelating data region by region
of the sky.
For the case analysed in this work, 
one can conclude that the MAXIMA data are compatible with
Gaussianity under the three goodness-of-fit methods applied.

\section*{Acknowledgments}

The authors would like to thank Julio Gallegos, Antonio Aliaga, 
Radek Stompor and Luis Tenorio 
for helpful comments.
LC, EMG, FA and JLS thank 
the Ministerio de Cienc\'\i a y Tecnolog\'\i a, projects ESP2001-4542-PE and
ESP2002-04141-C03-01.


\begin{thebibliography}{}


\bibitem{}Acquaviva, V., Bartolo, N., Matarrese, S. \& Riotto, A. 2002, astro-ph/0209156

\bibitem{}Aliaga, A.M., Mart\'\i nez-Gonz\'alez, E., Cay\'on, L., Arg\"ueso, F., Sanz, J.L. \& Barreiro, R.B. 2003, submitted to New Astronomy Reviews

\bibitem{}Balbi, A. et al. 2000, ApJ, 545, L1

\bibitem{}Bernardeau, F. \& Uzan, J-P. 2002, Physical Review D, 66, 103506

\bibitem{}Bromley, B.C. \& Tegmark, M. 1999, ApJ, 524, L79

\bibitem{}Cay\'on, L., Mart\'\i nez-Gonz\'alez, E., Arg\"ueso, F., Banday, A.J. \& Gorski, K.M. 2003, MNRAS, 339, 1189


\bibitem{}D'Agostino, R.B. \& Stephens, M.A. 1986, ``Goodness-of-fit techniques'', Marcel Dekker, INC. New York

\bibitem{}Durrer, R. 1999, New Astronomy Reviews, 43, 111


\bibitem{}Gangui, A., Lucchin, F., Matarrese, S. \& Mollerach, S. 1994, ApJ, 430, 447

\bibitem{}Gorski, K.M., Hivon, E. \& Wandelt, B.D. 1999, ``Evolution of
Large Scale Structure: from Recombination to Garching'', ed. A.J. Banday, 
R.K. Seth \& L.N. da Costa, astro-ph/9812350


\bibitem{}Gupta, S., Berera, A., Heavens, A.F. \& Matarresse, S. 2002, Physical Review D, 66, 043510

\bibitem{}Hanany, S. et al. 2000, ApJ, 545, L5


\bibitem{}Komatsu, E. \& Spergel, D.N. 2001, Physical Review D, 63, 063002 

\bibitem{} Komatsu, E., Wandelt, B.D., Spergel, D.N., Banday, A.J. \& 
Gorski, K.M. 2002, ApJ, 566, 19 


\bibitem{}Landriau, M. \& Shellard, E.P.S. 2002, astro-ph/0208540

\bibitem{}Mart\'\i nez-Gonz\'alez, E., Gallegos, J.E., Arg\"ueso, F., Cay\'on, L. \& Sanz, J.L. 2002, MNRAS, 336, 22 


\bibitem{}Peebles,P.J.E., 1999a, ApJ,510,523

\bibitem{}Peebles,P.J.E., 1999b, ApJ,510,531

\bibitem{}Rayner, J.C.W. \& Best, D.J. 1990, International Statistical Review, 58, 9


\bibitem{}Santos, M.G. et al. 2002a, Physical Review Letters, 88, 241302

\bibitem{}Santos, M.G. et al. 2002b, astro-ph/0211123

\bibitem{}Shapiro, S.S. \& Francia, R.S. 1972, Journal of American Statistical
Association, 67, 215

\bibitem{}Thomas, D.R.\& Pierce, D.A. 1979, Journal of the American Statistical Association, 74, 441 

\bibitem{}Wu, J.H.P. et al. 2001, Physical Review Letters, 87, 251303



\end{thebibliography}
\end{document}